# Optofluidic transport and particle trapping using an all-dielectric quasi-BIC metasurface

Sen Yang[2,1], Justus C. Ndukaife[3,2,1*]

[1]Vanderbilt Institute of Nanoscale Science and Engineering, Vanderbilt University, Nashville, TN, USA 37235

[2]Interdisciplinary Materials Science, Vanderbilt University, Nashville, TN, USA 37235

[3]Department of Electrical and Computer Engineering, Vanderbilt University, Nashville, TN, USA 37235

[*]*justus.ndukaife@vanderbilt.edu*

## Abstract

Manipulating fluids by light at the nanoscale has been a long-sought-after goal for lab-on-a-chip applications. Plasmonic heating has been demonstrated to control microfluidic dynamics due to the enhanced and confined light absorption from the intrinsic losses of metals. Dielectrics, counterpart of metals, is used to avoid undesired thermal effects due to its negligible light absorption. Here, we report an innovative optofluidic system that leverages a quasi-BIC driven all-dielectric metasurface to achieve nanoscale control of temperature and fluid motion. Our experiments show that suspended particles down to 200 nanometers can be rapidly aggregated to the center of the illuminated metasurface with a velocity of tens of micrometers per second, and up to millimeter-scale particle transport is demonstrated. The strong electromagnetic field enhancement of the quasi-BIC resonance can facilitate increasing the flow velocity up to 3-times compared with the off-resonant situation. We also experimentally investigate the dynamics of particle aggregation with respect to laser wavelength and power. A physical model is presented to elucidate the phenomena and surfactants are added to the particle colloid to validate the model. Our study demonstrates the application of the recently emerged all-dielectric thermonanophotonics in dealing with functional liquids and opens new frontiers in harnessing non-plasmonic nanophotonics to manipulate microfluidic dynamics. Moreover, the synergistic effects of optofluidics and high-Q all-dielectric nanostructures can hold enormous potential in high-sensitivity biosensing applications.

## Introduction

Controlling long-range transport of fluids has been a fundamental requirement in microfluidic systems, from standard flow cell assays to lab-on-a-chip devices. Traditionally, pressure-driven control technique and syringe-pumps have been widely used in microfluidics[1,2]. The evident scale mismatch between the microfluidic system and the bulky control system has inspired significant effort to develop integrated micrometer scale control techniques[3]. Among them, one promising approach to achieve integrated control of particle and fluid motion is to use light to control the flow of fluids, particularly at the micrometer scale, i.e., optofluidics[4–6]. The buoyancy-driven toroidal convection by heating the water with a laser beam can help transport and concentrate particles, but the flexibility to control the convection flow is limited[7–9]. One solution to achieve fluid manipulation is to use localized thermal gradients induced by light illumination[10].

Metal nanostructures, when illuminated by light at their plasmonic resonance, can tightly confine the energy in sub-wavelength scales in the vicinity of the structures[11]. The enhanced light absorption from the intrinsic (Ohmic) losses of metals has been considered as side effects such as limited quality factor (Q)[12,13], and reduced trapping stability in plasmonic nanowteezers due to undesired thermal heating effects[14,15]. Recently, however, scientists have realized that this enhanced light absorption can efficiently turn metal nanostructures into nanosources of heat, inspiring the study of thermoplasmonics[16–18]. This finding has found numerous applications in nanotechnology, namely, for photothermal cancer therapy[19], photothermal imaging[20], targeted drug delivery[21], solar-powered steam generation[22], as well as nanoscale control of temperature distribution[23] and thereby for optofluidics. Single[24] and arrays[25–27] of metal nanostructures have been studied to control the convection-driven dynamics. However, the surface of metal nanostructures may have unnecessarily high temperature which can be harmful to particles touching the surface[28].

All dielectric nanostructures, on the other hand, has been rapidly developing for the past decade mainly because they are relieved of parasitic Ohmic losses inherent by plasmonic nanostructures[29,30]. Many works have leveraged its low light absorption and thereby negligible Joule heating to avoid undesired fluid motion and/or positive (repulsive) thermophoresis which can deteriorate the optical trapping stability in plasmonic nanotweezers[31], such as by leveraging silicon dimers[32] and anapole-assisted nanoantennas[33–35]. Very recently, however, an emerged new field of all-dielectric thermonanophotonics focuses on controlling subwavelength optical heating by precisely tuning optical losses in dielectrics[36]. In ref [36], it was raised that temperature-gradient-driven microfluidic flows may also be controlled by dielectric nanoparticles. In this work, for the first time, we experimentally demonstrate the synergistic effects of optofluidic transport and particle trapping using an all-dielectric metasurface under the quasi-bound states in the continuum (quasi-BIC) resonance.

The concept of BIC, first proposed in quantum mechanics by von Neumann and Eugene Wigner in 1929[37], has rapidly emerged as a powerful approach for realizing high Q and strong field enhancement in dielectric metasurfaces[38]. The dielectric quasi-BIC metasurfaces are highly promising as they can have high-Q resonances comparable to photonic crystals[39], and strong field enhancements comparable to or even higher than those reported in plasmonic nanostructures. A variety of applications has been demonstrated in this field including lasing[40–42], biosensing[43–46], and nonlinear harmonic generation[47,48]. Among these designs, the intrinsic loss of materials has been identified as a critical issue for the quasi-BIC mode[49–51]. We reported that the loss of the surrounded environment (refractive index of water at 1.55 μm, $n$ = 1.31 + 0.00013i) can strongly affect the quasi-BIC resonance[52]. The absorptance from water in the proximity of the metasurface composed of elliptical silicon resonators can be as high as 45%, even though the loss from the resonators made by silicon is negligible. As shown by our previous work in which the water absorption in a single photonic crystal cavity served as a heat source for electrothermal effects[53], in this work, we instead discuss how to leverage the heat dissipation due to water absorption in a quasi-BIC system to engineer the microfluidic flow (see Fig. 1). Approaching the resonance, the total heat generation comprises of the global absorption by the bulk water in the microfluidic chamber as well as the heat dissipation from the water layer close to the resonators which serves as local heat sources. We show that this localized heating effect strongly increases the flow velocity up to about 3-times in comparison to outside the quasi-BIC resonators. The flow velocity can be precisely controlled by simply tuning the wavelength (within $\pm$7 nm range) and/or power of the input laser. Particles can be rapidly transported (with velocities of 10 to $10^2$ micrometer-per-second) from distances of up to millimeter-scale and aggregated at the center of the laser spot by the flow. The transported particles are then confined close to the metasurface by positive thermophoresis. Fluorophore-labeled tracer polystyrene (PS) beads with the size down to 200 nm are examined and we expect the same effects can be achieved for particles below 100 nm. Finally, we showcase totally different particle aggregation distributions when adding a cationic surfactant (cetyltrimethylammonium chloride) to the nanoparticle colloid by tuning the metasurface to on or off resonance. This verifies the temperature field distribution is modified by the quasi-BIC resonance. As such, our works opens a new field for controlling thermal-induced microfluidic dynamics at the nanoscale by non-plasmonic nanostructures.

# Main

Our system comprises an all-dielectric quasi-BIC metasurface sitting on the substrate of a microfluidic chamber illuminated with a collimated and linearly polarized laser beam, as shown in Fig. 1b. The laser spot follows Gaussian distribution, and the diameter is around 300 μm, after it is shrunken through a beam expander by 5-times. The laser power is amplified by an Erbium-doped fiber amplifier (EDFA) such that output of hundreds of milliwatts can be obtained. Relatively strong water absorption falls within the working range of the EDFA, with the absorption peak of water located at 1460 nm (see SI, Fig. S1). The metasurface is composed of elliptical silicon nanoantennas arranged in a zigzag array on a glass substrate (see Fig. 1a). When the input wavelength is away from the quasi-BIC resonance ($\lambda_{\mathrm{BIC}}$), the laser beam is transmitted after passing through the 160 μm height of water (i.e., chamber height). The absorption from water bulk yields to the Beer–Lambert law. This laser heating induces buoyancy-driven natural convection, which transports particles to the center of the illuminated substrate. Particles are confined close to the substrate in the axial direction (i.e., normal to the substrate) by the positive thermophoresis[54], as depicted in Fig. 1a. When the input wavelength is approaching $\lambda_{\mathrm{BIC}}$, the quasi-BIC resonance is excited. Highly localized field enhancement (see Fig. 2b) in the tip-to-tip gaps induces strong heat dissipation into water in the vicinity. These hotspots then serve as local heat sources inducing strong temperature gradient for manipulating microfluidic dynamics at the nanoscale. Besides the overall increased temperature which induces a faster buoyancy-driven flow, these local hot spots also induce strong thermos-osmotic flow due to the large lateral temperature gradient. Therefore, particles can be aggregated much more rapidly.

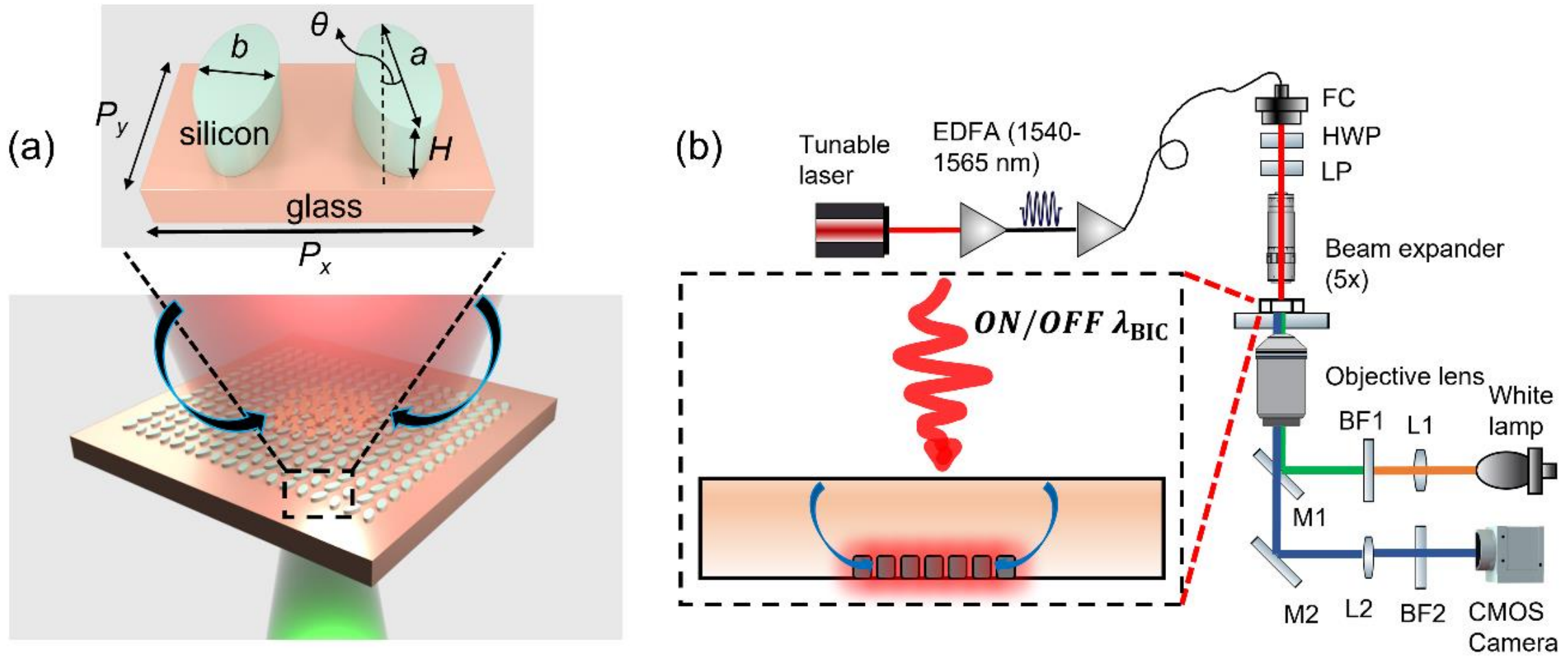

Figure 1. **Working principle and experimental facility. a** Schematic of the system. When the metasurface is off resonance, the laser heating of the bulk water induces buoyancy-driven flow, transporting and aggregating particles to the center of the illuminated region. When the quasi-BIC is excited, additional heat sources come from the heat dissipation of the water layer close to the resonators. The thermal-induced flow velocity is increased up to 3-times. The flow is represented by the two lobes near the nanoantennas. Inset: a unit cell of the metasurface. The geometrical parameters: periods, $P_x = 950$ nm, $P_y = 778$ nm; $a = 532$ nm, $b = 192$ nm, $H = 190$ nm, $\theta = 10°$. **b** Experimental set-up used for excitation of the quasi-BIC metasurface and imaging of the motion of suspended tracer particles. L1 and L2, focusing lenses; M1 and M2, mirrors; BF1 and BF2, bandpass filters used to filter light used for excitation of the fluorescent particles and light transmitted for imaging on the camera, respectively. Filtered fluorescent illumination are passed through the objective lens (10x or 40x) and focused to the sample. EDFA, Erbium-doped fiber amplifier used to amplify the power of the input laser; FC, fiber collimator; HWP, half wave-plate used to rotate the polarization direction of the laser beam; LP, linear polarizer. The metasurfaces and fluorescent tracer particles are visualized on a complementary metal-oxide-semiconductor camera by collecting signals through the same objective lens.

The simulated and measured spectra of the metasurface are shown in Fig. 2a, respectively. For an infinite array, no light is transmitted and the absorption of water (red dash line in Fig. 2a) accounts for the 10% drop in the reflected power. In our experiments, each metasurface is a 500 μm by 500 μm square array, large enough compared with the laser spot diameter. The finite array size and the fabrication uncertainties (see Fig. 2c for the scanning electron microscopy image) suppress the performance of the resonance mode and increase the background noise during measurement. Therefore, a higher transmittance (from 0% to 46.1%) and a reduced Q (from 500 to 250) are observed in the measured spectra. We expect this suppressed performance impairs the heating effect while our experiments show that it is still strong enough to modify the whole temperature field distribution (see experiment results in Fig. 3 and 4). The large near-field electric field enhancement close to the surface of the nanoantennas contributes to such strong heating effects, as shown in Fig. 2b. It is worth noting that the BIC mode we employ is a symmetry-protected mode[55], which is sensitive to the incident angle and the array size[56]. This is the key reason we choose to use a collimated laser beam instead of a focused laser beam by objective lens. Improved control flexibility with smaller illuminated areas and lower incident power can be achieved by leveraging resonance-trapped BICs[57] and/or merging BICs[39,58].

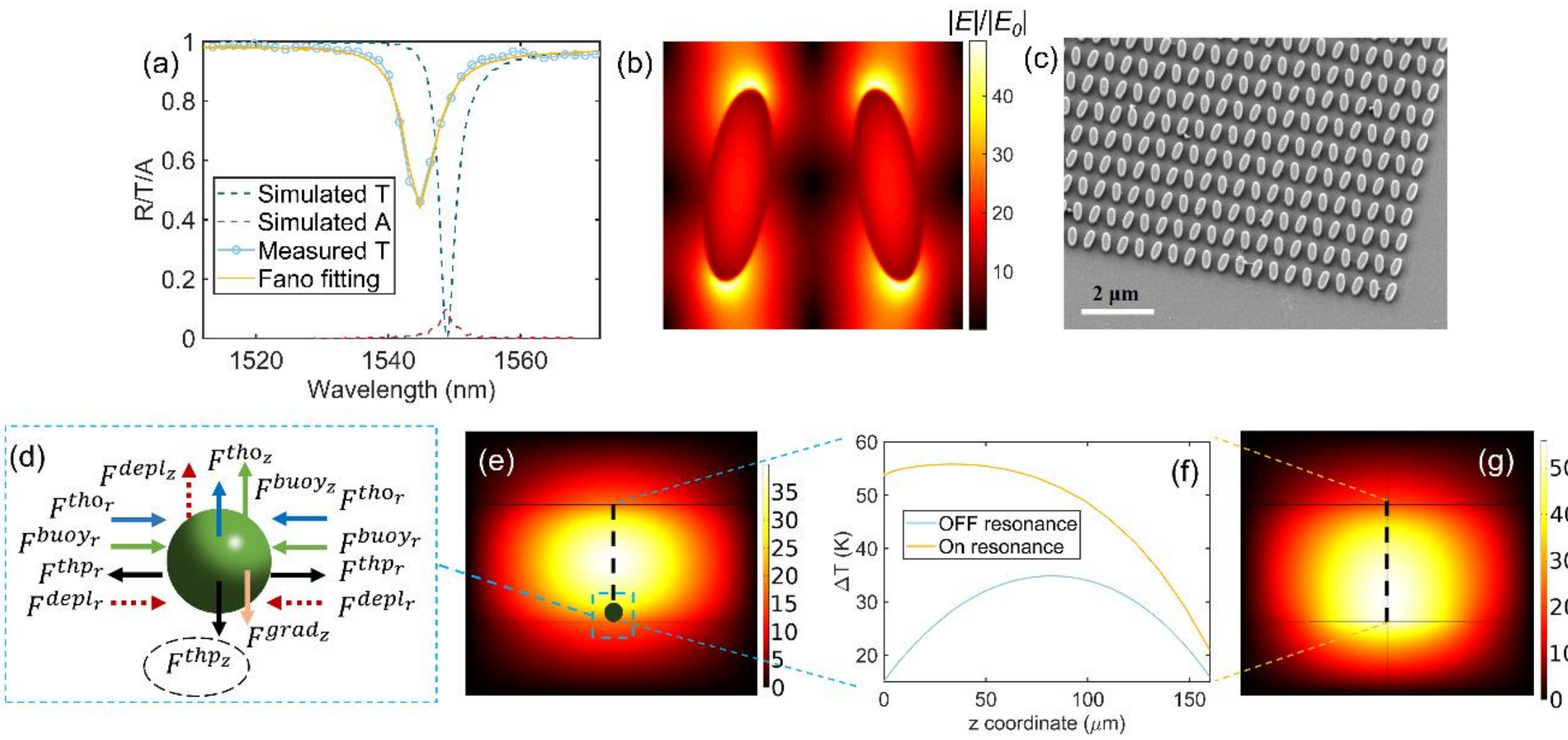

Figure 2. **Optical characterization and thermal simulation**. **a** Simulated and measured spectra of the metasurface. The resonance positions are 1548.9 nm and 1544.8 nm for the simulation and measurement, respectively. The sample shows a larger transmittance (46.1%) and a larger linewidth (6.3 nm) compared to the simulations (0% and 3.0 nm, respectively) attributed to the finite size and fabrication imperfections. The well overlapped Fano fitting curve (yellow solid line) with the measured transmittance spectrum validates a typical quasi-BIC resonance. **b** Electric field enhancement distribution within one unit cell. The maximum electric field enhancement factor is 49.4, i.e., 2440-times for the intensity enhancement, supporting strongly enhanced water absorption. **c** Representative scanning electron microscopy image of the fabricated metasurface. The array size is 500 μm. **d** Depiction of the major forces acting on the particles. tho, thermos-osmosis; buoy, buoyancy; thp, thermophoresis; grad, optical gradient; depl, depletion-attraction. **e** Simulated temperature field distribution for off-resonant and **g** on-resonant conditions. Black dash lines mark the lines from which the spatial temperature rise curves in f are extracted. **f** Temperature rise in *z* direction for off-resonant and on-resonant conditions. *z* = 0 is the surface of the glass substrate.

To understand the modulation of the heat distribution by the quasi-BIC resonance, we numerically solved the temperature filed by a commercially available finite element method software package (COMSOL Multiphysics). The temperature field in the system is determined by solving the steady-state heat equation given by

$$\nabla \cdot \left[\kappa \nabla T(\mathbf{r}) + \rho c_{\mathrm{p}} T(\mathbf{r}) \mathbf{u}(\mathbf{r})\right] = q(\mathbf{r}) \qquad (1)$$

The first term on the left corresponds to the heat conduction, while the second term corresponds to the convection depending on the velocity of the flow. $T(\mathbf{r})$ and $\mathbf{u}(\mathbf{r})$ are the spatial temperature and fluid velocity field, respectively; $\kappa$, $\rho$ and $c_p$ are thermal conductivity, density and heat capacity of the materials, respectively; and $q(\mathbf{r})$ is the heat source density giving the heat dissipated per unit volume. In our system, $q(\mathbf{r})$ comprises of two parts. The first part comes from the light absorbed by the bulk water modelled by[25]

$$q(\boldsymbol{r}) = P_0 \frac{\alpha_{\mathrm{c}}}{2\pi\sigma^2} \mathrm{e}^{-\left(\frac{\mathrm{r}^2}{2\sigma^2}\right)} \mathrm{e}^{-\alpha_{\mathrm{c}} \mathrm{z}} \qquad (2)$$

$$\alpha_{\mathrm{c}} = \frac{4\pi\bar{\kappa}}{\lambda} \qquad (3)$$

where $P_0$ is incident laser power, $\lambda$ is laser wavelength, $\alpha_c$ is the attenuation coefficient of water, $\bar{\kappa}$ is the imaginary component of the refractive index of water. σ is defined by $H = 2(2\ln 2)^{1/2}\sigma$ which is the fwhm of the beam. The second part comes from the light absorbed by water near the nanoantennas (hotspots) when the quasi-BIC resonance is excited. For computational simplicity, we approximate the heat dissipation obtained from electromagnetic calculations as uniform at each hotspot while taking into account the spatial Gaussian distribution of the incident laser beam (see S3 of SI for details).

To better compare the heat distributions when on and off resonance, we simulate the temperature fields in the two cases, as shown in Fig. 2e and 2g, respectively. The temperature drops from the middle of the chamber down to the substrate. Therefore, particles experience a positive thermophoretic force[64,65] which pushes particles towards the substrate and acts as the main mechanism to confine particles in the axial direction. Since the thermophoretic force is proportional to $S_T \nabla T$ where $S_T$ is the Soret coefficient, the confinement stability depends on the temperature gradient in the axial direction. Although the highest temperature rise increases from 38 K to 56 K, the hot center is brought down from chamber center in the off-resonant condition to the vicinity of the substrate in the on-resonant condition due to the strong heating effects from the resonance, as depicted in Fig. 2f. Therefore, the temperature gradient in the axial direction is decreased significantly in the on-resonant condition, leading to decrease of the downward thermophoretic force. Meanwhile, the stronger heating induces a stronger flow which exerts stronger drag forces on particles to take them away from the substrate. Thus, particle aggregation will become less stable in the axial direction when too close to the resonance, as presented in Fig. 4c.

Based on the temperature field analysis, the key forces acting on the trapped particles are depicted in Fig. 2d. The drag force $F^{buoy}$ from the buoyancy-driven convection and $F^{tho}$ from the thermo-osmotic flow bring particles towards the center in the lateral direction and take particles away in the axial direction. The positive thermophoretic force $F^{thp}$ pushes particles from the middle down towards the substrate in the axial direction. When the quasi-BIC mode is excited, another force that pulls particles to the hotspots is the optical gradient force $F^{grad}$ derived from the highly enhanced electric field, which is short-range and close to the nanoantennas. Due to the strong heating effects in our system, we consider the downward $F^{thp}$ is the main contribution to confine aggregated particles close to the substrate.

The representative aggregation of particles when illuminating the metasurface is shown in Fig. 3a. We used 500 nm fluorophore-labeled PS beads as tracer particles to visualize the microfluidic flow. Experimental demonstrations using 200 nm PS beads are shown in S2 of the supplementary videos. The experimental videos are processed using an open-source particle tracking analysis package called *Trackpy*[59,60]. Representative particle trajectories are presented in Fig. 3b. It can be found that particles are rapidly transported towards the laser spot center and aggregated close to the substrate. The empty region at the center of the trajectory map represents a stagnation zone, i.e., the aggregated particle cluster. The experimentally measured angularly averaged radial velocities obtained from the particle tracking analysis are shown in Fig. 3c and 3d. As shown in Fig. 3c, the fluid radial velocity when the collimated laser beam is illuminated on the metasurface increases rapidly as the input wavelength approaches the resonance (1544.3 nm). The results show a maximum flow radial velocity of 45 μm/s at 1545.3 nm, two times larger than that at 1551.3 nm. Thus, the flow velocity can be precisely controlled in a wide range by simply tuning the wavelength within ±7 nm band which profits from the high Q attribute of the quasi-BIC resonance. This is not achievable in plasmonic arrays. For example, in ref [26], the wavelength distance for the 2-times difference of the maximum velocity is almost 100 nm, two orders of magnitude larger compared with our case. The radial flow velocity close to the metasurface initially increases from a radial distance of more than ~550 μm (limited by the field of view) until it reaches its maximum at a radial distance of ∼180 μm from the particle cluster. Inward from this position, the radial velocity decreases towards the center of the laser spot. Thus, we expect particles within at least a range of 600 micrometer-scale radius can be rapidly captured and transported to the center of the illuminated region. The scaling of the flow velocity with laser power is shown in Fig. 3d. The flow velocity rises when the input power is increased. The particle tracking analysis is not applicable when the input power is larger than 500 mW as the particles move too fast to be accurately located.

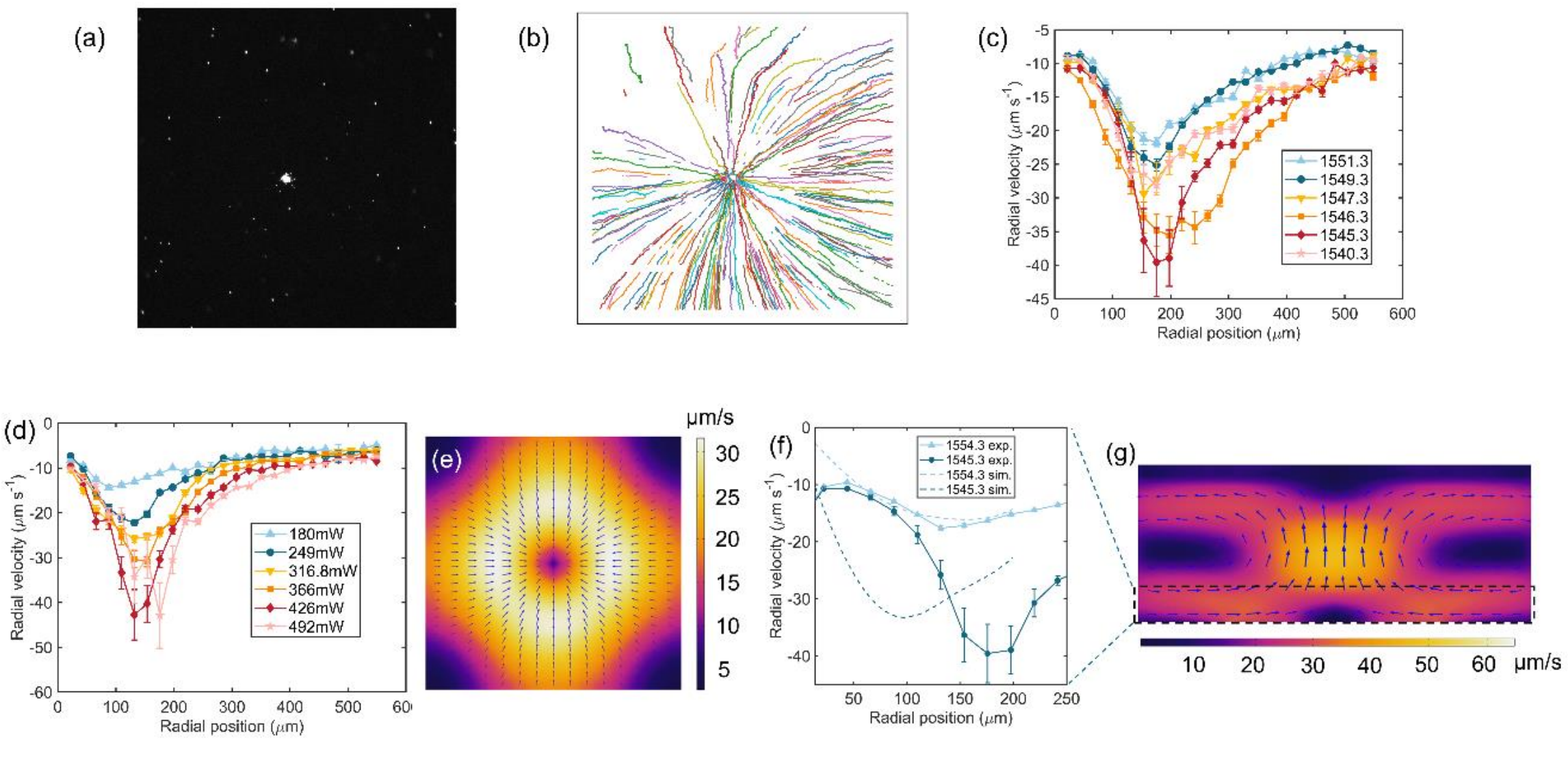

Figure 3. **Experimental and simulation results for particle transport.** All data are obtained under a 10x objective lens. The filed of view is 900 μm. **a** Representative particle aggregation when collimated laser beam is illuminated on the metasurface. **b** Representative particle trajectory map extracted from sequential 600 frames of a recorded video. The frame rate is 10 frames per second. The empty region at the center indicates that particle movement is trivial in this area, corresponding to the aggregated particle cluster. The result shows that the flow is directed radially inwards towards the center of the laser spot and serves as a powerful means to concentrate suspended particles to the vincinity of the metasurface. It's noted that cut-offs can be observed for some particle trajectories. This is due to the low transmittance of silicon in the visible range, which dims the fluorescence of these tracer particles, making them hard to be tracked. **c** Scaling of experimentally measured radial flow velocity with laser wavelength. The negative sign

represents inward direction. The error bar shows the standard error of the mean. The laser power fluctuates around 420 mW. The position where velocity reaches maximum is slightly farther away from the center for wavelenths closer to the resonance. We attribute this to the stronger positive (respusive) thermophoresis in the lateral direction due to stronger heating effect. **d** Scaling of experimentally measured radial flow velocity with laser power. The same repulsive phenomenon is observed for higher laser power. **e** Simulated flow velocity distribution in the *xy*-plane (5 μm above the substrate) and **g** in the *xz*-plane. Superimposed arrows show the direction of the flow vectors. Velocities in the black dash box in c are averaged to obtain the dash lines in b. **f** Simulated (dash lines) and measured (solid lines) radial flow velocity for near-resonance (1545.3 nm) and off-resonance (1554.3 nm). The maximum velocities for 1545.3 nm and 1554.3 nm are 45 μm/s and 17 μm/s, respectively.

To understand the physics of the observed thermal-induced microfluidic dynamics in this system, we numerically solved the flow field by COMSOL Multiphysics. We consider that two mechanisms contribute to the thermal-induced flows in this system: the buoyancy-driven convection and thermo-osmotic flow. The velocity field distribution is determined from the solution of the incompressible Navier–Stokes equation given by[61]

$$\rho_0(\mathbf{u}(\mathbf{r})\cdot\nabla)\mathbf{u}(\mathbf{r})+\nabla p(\mathbf{r})-\eta\nabla^2\mathbf{u}(\mathbf{r})=\mathbf{F} \qquad (4)$$

where $\nabla\cdot\mathbf{u}=0$, $\rho_0$, $p(\mathbf{r})$ and $\eta$ are fluid density, pressure, and dynamic viscosity, respectively; and $\mathbf{F}$ is the force per unit volume acting on the fluid element. For the buoyancy-driven convection, we employ the Boussinesq approximation and gives[24,26]

$$\boldsymbol{F}_{\text{buoy}} = g\rho_0\beta(T)[T(\boldsymbol{r})-T_0] \qquad (5)$$

where $g$, $\beta(T)$ are the gravitational constant, and thermal expansion coefficient of water, respectively. For the thermos-osmotic flow, we apply a slip velocity to the thin fluid layer close to the surface of the substrate given by[62,63]

$$v_{\parallel} = \chi\frac{\nabla_{\parallel}T}{T} \qquad (6)$$

where $\chi$ is thermo-osmotic coefficient dependent on the properties of the liquid–solid interface (ζ potential, Hamaker constant, etc.) and $\nabla_{\parallel}T$ is the temperature gradient parallel to the surface.

The simulated velocity distributions in the *xy*-plane near the substrate and in the *xz*-plane are shown in Fig. 3e and 3g, respectively. The flow close to the substrate brings particles inwards towards the center of the illuminated area and drags particles away from the surface in the axial direction (*z* direction). Due to the large focal depth of the 10x objective lens, the measured radial velocities are averaged from particles that appear within 30 μm from the surface (see Fig. S2 for details). Similar simulation results from such averaging are compared with experiment results in Fig. 3f. The measured maximum flow velocity at 1545.3 nm reaches around 3-times of that at 1554.3 nm (11 nm away from resonance). For the off-resonant condition, the simulation agrees well with the measurement. For the near-resonant condition, the trend of the simulation remains consistent with the measurement despite the underestimation. We attribute this difference mainly to the assumption of uniform heat distributions at each resonant nanoantenna. Our simulations show that the thermos-osmotic flow is weak in the off-resonant condition and it is insensitive to the changes of the thermo-osmotic coefficient $\chi$. However, when the quasi-BIC mode is excited, the thermos-osmotic flow becomes an important contribution to the total transport flow. This is due to the strong $\nabla_{\parallel}T$ provided by the hotspots near the nanoantennas. As shown in Fig. 2b, the electric field distribution in the tip-to-tip gaps is not uniform, and thus should induce a stronger $\nabla_{\parallel}T$ and consequently a stronger thermos-osmotic flow. Besides, our estimation for the zeta potential, ζ, of the water-substrate may also play a part.

We have demonstrated the thermal-induced flow as a powerful tool for controling long-range particle transport. Next, we discuss the trapping performance of our system. We compare the evolution of particle aggregation with illumination time when the input wavelength is near-resonant (1545.3 nm) and off-resonant (1551.3 nm), as shown in Fig. 4a and 4b. It can be found that particles are concentrated in a faster manner for the near-resonant illumination. More specifically, the time to concentrate particles to reach a similar number of aggregrated particles in the near-resonant condition is only two-thirds of that in the off-resonant condition. This agrees with the aforementioned flow velocity measurements. Moreover, the particle cluster is packed more tightly in the near-resonant condition as shown in the last frame of Fig. 4a and 4b.We attribute this to the stronger thermal-induced flow directing towards the center due to stronger heating effects, which exerts larger drag forces on the particles. Notably, approaching the resonance condition does not always improve the particle aggregation. The particle cluster at different input wavelength is shown in Fig. 4c. The cluster is packed more tighly from 1553.3 nm to 1546.3 nm as approaching the resonance, shown by the effective particle distribution region (see S1 of SI) marked by the red circles. However, particles start to be lost when the input wavelength is closer to resonance (1545.3 nm) in which the number of particles are decreased, while the cluster packing is even more tightly. On resonant condition (1544.3 nm), particles can not be aggregrated any more. The slightly different resonance position from the spectra measurement (Fig. 2a) comes from the sample loading process as the symmetry-BIC mode is sensitive to the incident angle. Three nanometers away from resonance (1541.3 nm), the particle cluster appears again. The mechanism for these phenomena are two fold. First, at resonance (or very close to resonance), the temperature gradient in the axial direction is no longer strong enough to stably confine particles to the substrate by the positive thermophoresis. Second, the thermal-induced flow is stronger, resulting in stronger drag forces in the axial direction to take particles away from the substrate. We anticipate this can be improved by supressing the quasi-BIC mode properly to reduce the heating effects from the high electric field enhancement, for example, a larger tilt angle $\theta$ for our case.

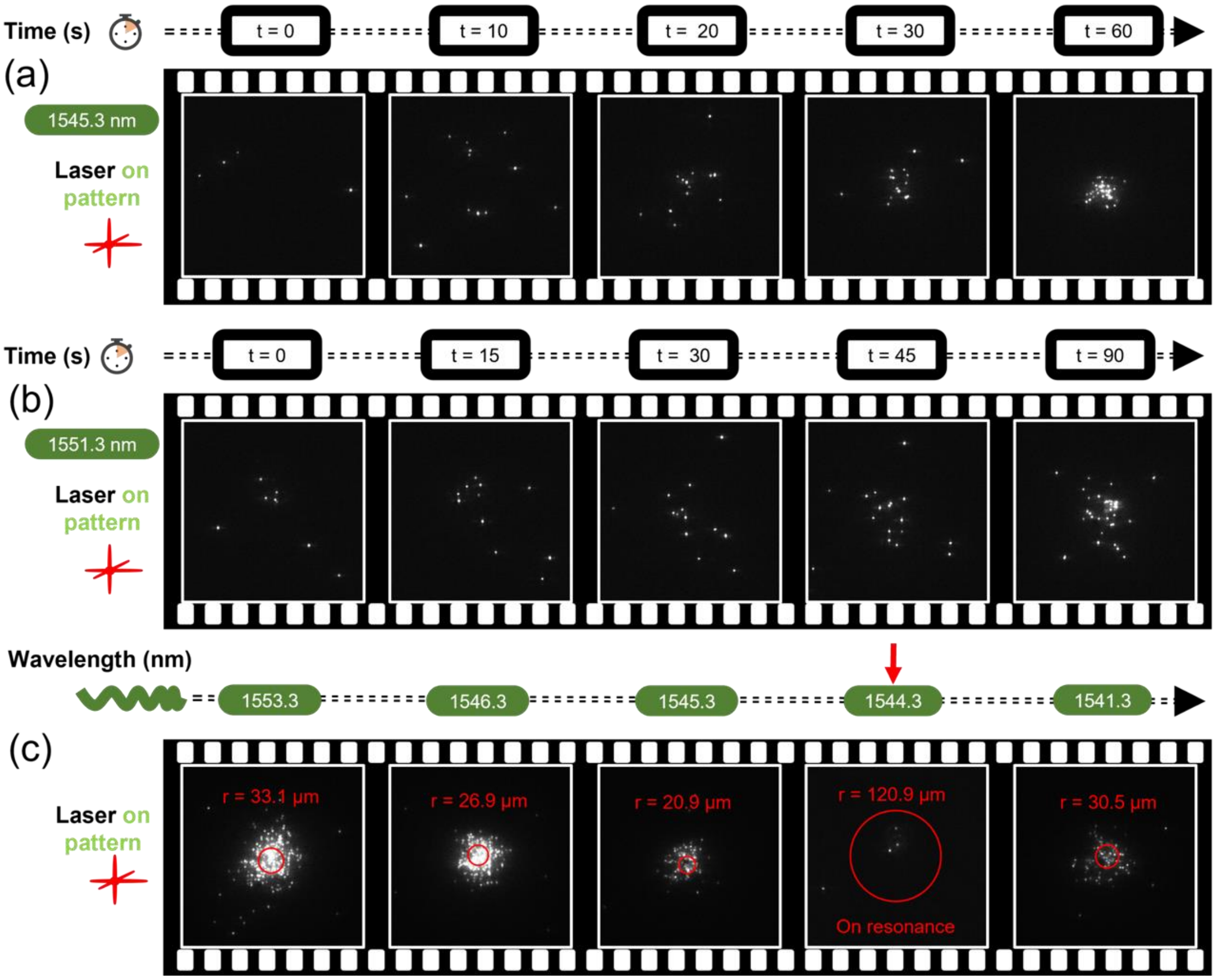

Figure 4. **Experiment results for particle aggregation.** All data are obtained under a 40x objective lens. The laser power fluctuates aroung 270 mW. **a** Evolution of particle aggregation with illumination time when the input wavelength is near resonance (1545.3 nm). **b** Evolution of particle aggregation with illumination time when the input wavelength is off resonance (1551.3 nm). Particles are aggregrated more rapidly and packed more tightly in near-resonant condition. **c** Evolution of particle aggregation with illumination wavelength. The cluster is packed more tightly when approaching resonance, while starts to lose particles when very close to resonance. No particles can be aggregrated on resonance. Particles are aggregrated again when away from resonance.

Finally, to validate our analysis, we added a cationic surfactant, (cetyltrimethylam-monium chloride, CTAC) to the nanoparticle colloid to experimentally investigate the temperature distribution when on and off resonances. CTAC has been widely used for reversing the sign of the effective thermophoresis, i.e., inducing negative (attractive) thermophoresis to attract particles towards hot regions[66–68]. CTAC molecules adsorbed on the PS particle surface can form a positively charged molecular double layer. Simultaneously, CTAC molecules self-assemble into micelles when above the critical micelle concentration (0.13–0.16 mM). As reported in ref [67], when the CTAC concentration is below ~2 mM, a thermoelectric field is generated to attract particles towards the hot region. While when the CTAC concentration is above ~2 mM, the depletion of the CTAC micelles plays the key role in attracting particles, i.e., the depletion-attraction force (DAF). In brief, DAF is an osmotic pressure exerted on particles to attract them towards hot regions due to the concentration gradient of micelles generated from the migration of micelles from hot to cold regions[69]. In our case, we consider the later mechanism as the CTAC concentration in our experiments is 5 mM.

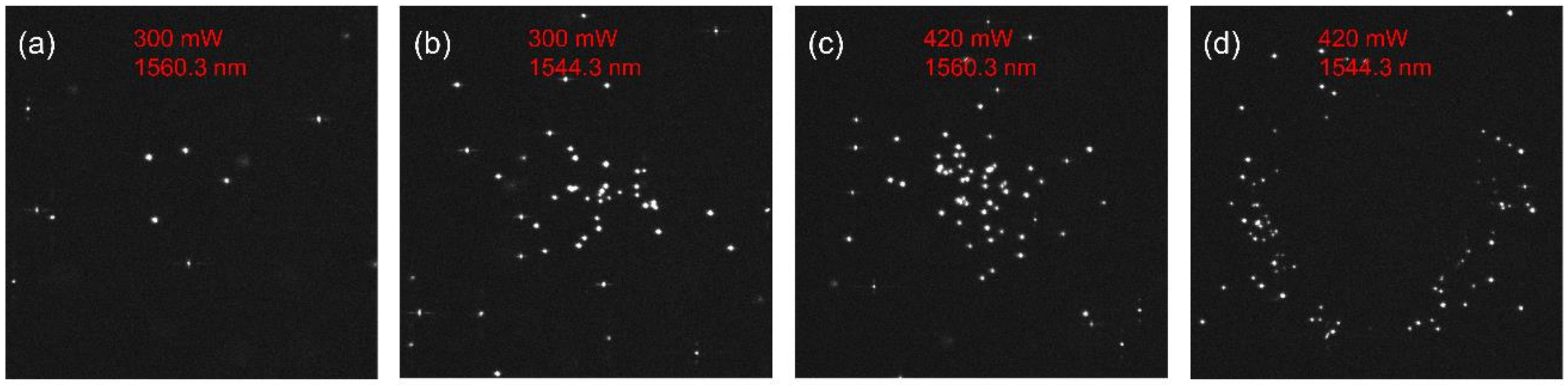

Figure 5. **Experiment results for CTAC solution.** CTAC concentration is 5 mM. All data are obtained under a 10x objective lens. **a** Particles are hard to accumulate at low power (300 mW) and **c** aggregated at high power (420 mW) for the off-resonant condition (1560.3 nm). **b** Particles are aggregated at low power (300 mW) and **d** localized as a ring at high power (420 mW) for the on-resonant condition (1544.3 nm). The aggregation is not as tight as that in Fig. 2a mainly due to the stronger thermophoresis repelling particles from the center.

Fig. 5 shows totally different particle distributions for on and off resonance cases, respectively. This verifies our aforementioned discussions that the temperature field distribution is modified by the quasi-BIC resonance of the nanoantennas. For the off-resonant condition, particles are hard to be aggregated on the substrate at low laser power (Fig. 5a) as they are attracted towards the hot center at the center of the chamber (see Fig. 2d and 2e) by the DAF. At high power (Fig. 5c), however, particles are aggregated again as the positive thermophoresis overcomes the DAF to push particles down. On the other hand, for the on-resonant condition, since the hot center is close to the substrate, particles are aggregated at low power (Fig. 5b) but show a ring-like localization at high power (Fig. 5d). This transition from accumulation to ring-like distribution is due to the interplay between thermophoresis and DAF and has been reported by several research groups where the hot spot is located at the top wall of the chamber and the concentration of the added polymers is varied[69–71]. In these works, the chamber height is usually less than 10 μm to suppress thermal convection. In a very recent work, David Simon et al. reported that the transition can also happen between low and high laser power[72]. We emphasize that the positive thermophoretic force in this experiment is strong enough to repel particles in the lateral direction to generate the ring-like distribution. This enhanced positive thermophoresis is attributed to the following reason. The magnitude and even the sign of the $\zeta$ potential of the particles can be significantly modified by the adsorbed surfactant molecules[73]. In our experiments, the $\zeta$ potential of the 500 nm PS beads is modified from -40 mV suspended in deionized water to +88 mV with the CTAC concentration of 5 mM (see Fig. S3 of SI). If we neglect the permittivity and salinity gradients, the thermophoretic mobility[62,73] of the particle is proportional to $\zeta^2$. In this case, the natural thermophoretic force repelling particles from the center increases up to around 5-times. We expect this ring-like distribution can be beneficial for trapping particles at cool regions as well as separating different particles[71].

# Discussions

Besides manipulating microfluidic dynamics, we expect the system we have proposed combining optofluidics and high-Q all-dielectric nanostructures can offer significant potential for biosensing applications. Compared with dielectrics, the nonradiative absorption resulting from the Ohmic losses of metals naturally limits the Q of the resonance, which is crucial for ultrasensitive sensing[74] and lasing applications[75]. Hatice Altug et al. have demonstrated several works employing all-dielectric quasi-BIC metasurfaces as label-free nanophotonic biosensors[43–46], which have presented promising sensor performances. However, in these systems, the analytes are immobilized on the sensor surface through either dropping droplets, spin coating, or pressure-driven flow. We expect that the ability of our system to concentrate particles suspended in the liquid can further push down the detection limit of such quasi-BIC-based biosensors and in situ refractometric detection is also possible. What's more, due to the strong near-field enhancement, quasi-BIC metasurfaces have been demonstrated to boost surface-enhanced Raman and fluorescence spectroscopy[76,77]. Benefiting from the more broadly spectral operation range of resonant dielectric nanostructures[74], this system can well separate the operation wavelengths between the optofluidic transport and the high-sensitivity particle detection. Therefore, this system may provide high signal-to-noise ratio as well as quenching-free fluorescence spectroscopy which is a fundamental limit in plasmonics[78].

In summary, we have introduced and demonstrated the nanoscale control of temperature and fluid motion in an all-dielectric system. Thanks to the high Q and strong electromagnetic field enhancement of the quasi-BIC mode, we present precise control of the fluid velocity up to 3-times by simply tunning laser wavelength within several nanometers. We also show long-range (millimeter-scale) and rapid (tens of micrometer per second) particle transport and aggregation. The undesired reduction of trapping stability at resonance is observed attributed to the modified temperature field which is altered by the strong heating effects from resonant nanoantennas. This can be improved by slightly suppressing the quasi-BIC mode, for example by illumination with the wavelength slightly away from the resonance or increasing the tilt angle of the nanoantennas in our case. By implementing a physical model, we numerically show how the quasi-BIC resonance altered the temperature field and fluidic dynamics at the nanoscale. Moreover, after adding a cationic surfactant, CTAC, to the nanoparticle colloid, the totally different particle distributions in on and off resonant conditions further validate our model. The ring-like particle localization arisen on resonance also provides potential applications in low-temperature particle trapping and particle separation.

In addition, we propose that, this system can become a powerful tool in colloid science and life science. Although we show aggregation of polystyrene beads down to 200 nm, we expect the same effect can have generality in particle size, shape, and composition, for example for sub-100nm nanoparticles. This can offer many opportunities in biology and medicine such as for concentration and detection of extracellular vesicles and viruses. What's more, as we have discussed, the combination of optofluidics and high-Q all-dielectric nanostructures can have great potential in boosting the sensitivity of biosensors benefitting from synergistic effects of effective particle aggregation and strong electromagnetic field enhancement in the near field.